\documentclass[a4paper, 11pt]{article}
\usepackage{amsmath, amssymb, amsthm}
\usepackage{graphicx}
\usepackage{latexsym}

\def\b{\begin{eqnarray}}
\def\e{\end{eqnarray}}
\def\n{\noindent}
\numberwithin{equation}{section}

\begin{document}

\begin{center}

{\Large\textbf{Algebraic Discretization of the Camassa-Holm and
Hunter-Saxton Equations}}

\vspace {5mm}

\noindent {\bf Rossen I. Ivanov}\footnote{Present address: School
of Mathematical Sciences, Dublin Institute of Technology, Kevin
Street, Dublin 8, Ireland, Email: rivanov@dit.ie}

\vskip.5cm

\begin{tabular}{c}
\hskip-1cm {\it Department of Mathematics, Lund University, 22100
Lund, Sweden}

\hskip-.8cm
\end{tabular}
\vskip1cm
\end{center}

\begin{abstract}

\noindent The Camassa-Holm (CH) and Hunter-Saxton (HS) equations
have an interpretation as geodesic flow equations on the group of
diffeomorphisms, preserving the $H^1$ and $\dot{H}^1$
right-invariant metrics correspondingly. There is an analogy to
the Euler equations in hydrodynamics, which describe geodesic flow
for a right-invariant metric on the infinite-dimensional group of
diffeomorphisms preserving the volume element of the domain of
fluid flow and to the Euler equations of rigid body whith a fixed
point, describing geodesics for a left-invariant metric on SO(3).
The CH and HS equations are integrable bi-hamiltonian equations
and one of their Hamiltonian structures is associated to the
Virasoro algebra. The parallel with the integrable SO(3) top is
made explicit by a discretization of both equation based on
Fourier modes expansion. The obtained equations represent
integrable tops with infinitely many momentum components.

An emphasis is given on the structure of the phase space of these
equations, the momentum map and the space of canonical variables.
\end{abstract}

\section{Introduction}

This geometric interpretation of the Camassa-Holm equation
\cite{CH93} as a geodesic flow equation on the group of
diffeomorphisms, preserving the $H^1$ right-invariant metrics
metric was noticed firstly by Misio\l ek \cite{M98} and developed
further in many recent publications, e.g. \cite{KM03,HMR98,
CK03,CK06,K04,CI07}. The CH equation has an interpretation in the
context of water waves propagation
\cite{CH93,J02,J03,DGH03,DGH04,I07,I-K07}. The spectral problem
for the CH equation on the line is developed in \cite{BBW07,
C01,CGI,CGI2,dMS,K06}, the periodic spectral problem -- in
\cite{C98,CM99,V07}. The CH solutions are investigated in a
variety of recent papers, e.g. in
\cite{BC07,BC07a,CE98,CI06,DM90,EY08,GKKV08,H05,PLA05,QZ06}.
Hierarchies of CH equations are studied in
\cite{CGI2,I06Z,I07DCDS}, different modifications are studied in
\cite{LQ07,SS07}.

There are different forms of the CH equation, containing linear
term with a first derivative $u_x$; with a third derivative
$u_{xxx}$ (called sometimes Dullin-Gottwald-Holm equation
\cite{DGH03,DGH04,M06,M07,Z07}), or without such terms. These
terms can be put in or removed from the equation independently by
Galilean transformations.

We will be interested in the CH equation of the form \b
m_t+au_{xxx}+2mu_x+m_xu=0,\qquad m=u-u_{xx}, \label {CH_1}\e

\n with $a$ being an arbitrary constant. It can be written in
Hamiltonian form \b m_t&=&\{m,H_1\}, \label{CHPB}\e

\n where, assuming that $m$ is $2\pi$ periodic in $x$, i.e.
$m(x)=m(x+2\pi)$, the Poisson bracket and the Hamiltonian are \b
\{F,G\}&\equiv& -\int_{0} ^{2 \pi}\frac{\delta F}{\delta m}\Big(
a\partial^3+ m \partial+\partial\circ m
\Big)\frac{\delta G}{\delta m}{\text d}x, \label{PB_1}\\
H_1&=&\frac{1}{2}\int_{0} ^{2 \pi} mu {\text d}x. \label{H_1} \e

\n The equation (\ref{CH_1}) is bi-Hamiltonian with a second
Hamiltonian representation $ m_t=\{m,H_2\}_2$, where \b \{F,G\}_2
&\equiv& -\int_{0} ^{2 \pi}\frac{\delta F}{\delta
m}( \partial-\partial^3)\frac{\delta G}{\delta m}{\text d}x, \label{PB_2}\\
H_2&=&\frac{1}{2}\int_{0} ^{2 \pi} (u^3+uu_x^2-\frac{a}{2}u_x^2 )
{\text d}x. \label{H_2} \e

\n One can notice that the integral \b H_0=\int_{0} ^{2
\pi}m{\text d}x \label{Cas}\e is a Casimir for the second Poisson
bracket (\ref{PB_2}).

The relation of the first Poisson bracket (\ref{PB_1}) to the
Virasoro algebra can be seen as follows \cite{D03}. The
$2\pi$-periodic function allows a Fourier decomposition \b
m(x,t)&=&\frac{1}{2\pi}\sum_{n\in\mathbb{Z}}L_n(t)
e^{inx}+\frac{a}{2}, \label{exp-m} \e

\n (the reality of $m$ can be achieved by $L_{-n}=\bar{L}_n$).
Then the Fourier coefficients $L_{n}$ close a classical Virasoro
algebra of central charge $c=-24\pi a$ with respect to the Poisson
bracket (\ref{PB_1}): \b \label{eq8}
i\{L_{n},L_{m}\}=(n-m)L_{n+m}-2\pi a(n^{3}-n)\delta_{n+m,0}.\e

\n The CH equation in the form \b m_t+2\omega
u_{x}+2mu_x+m_xu=0,\qquad m=u-u_{xx},\label{CH}\e

\n can be obtained from (\ref{CH_1}) via $u\rightarrow u+a$, and
apparently $\omega=3a/2$.

%or (\ref{eq7}) we have \b \label{equ} u(x)=\sum_{n\in
%Z}\frac{L_{n}}{1+n^2}e^{inx}+\frac{c}{24},\e

% and therefore, from (\ref{H_1}) \b \label{H_top} H_1=
%\pi\sum_{n\in Z}\frac{L_{n}L_{-n}}{1+n^2} +\frac{\pi c}{6}L_0.\e

\n Since \b H_0=\frac{L_0}{2\pi}+\pi a  \label{H0}\e is an
integral of motion (Casimir), so is $L_0$.

The first Hamiltonian is \begin{equation}  H_1=
\frac{1}{4\pi}\sum_{n\in \mathbb{Z}}\frac{L_{n}L_{-n}}{1+n^2}
+\frac{a}{2}L_0+\frac{2\pi a^2}{8}.\label{ham} \end{equation}

From (\ref{eq8}) and (\ref{ham}) we obtain the 'Camassa-Holm top'
equations on the Virasoro group, which are a discretization of the
Camassa-Holm equation (\ref{CH_1}) \b
i\dot{L}_{k}=\frac{1}{2\pi}\sum_{n\in
\mathbb{Z}}\frac{k+n}{1+n^2}L_{n}L_{k-n} +\frac{a
}{2}\frac{3k-k^3}{1+k^2}L_k ,\label{VT} \e

\n (the dot is a $t$-derivative). This equation is analogous to
the Euler top (rigid body) equation on the Lie group SO(3) \b
\nonumber \dot{M}_k=\sum_{p,l=1}^{3}\varepsilon_{kpl}\Omega_p M_l,
\qquad M_k\equiv I_k \Omega_k\e for the quadratic Hamiltonian \b
H_E=\frac{1}{2}\sum_{p=1}^3M_p\Omega_p ,\nonumber \e

\n where $I_k$ ($k=1,2,3$) are three constants -- the principle
inertia momenta. The phase space is embedded in the Lie coalgebra
so(3)* as a coadjoint orbit. The Lie-Poisson bracket, related to
the so(3)* coalgebra is \b \label{so(3)}
\{M_{n},M_{m}\}=\varepsilon_{nmk}M_k.\e

\n The inertia operator $\mathbf{I}$: so(3)$\rightarrow$ so(3)*
(see e.g. \cite{A}) relates the parametrization on the so(3)
algebra given by the functions $\Omega_k$ and the parametrization
on the co-algebra so(3)* given by the functions $M_k=I_k\Omega_k$.
Note that the Poisson bracket (\ref{so(3)}) has a Casimir \b
K=\Omega_{1}^2+\Omega_{2}^2+\Omega_{3}^2, \label{K}\e constraining
the phase space on a sphere. Since the Lie-Poisson bracket is
degenerate on so(3)*, the coadjoint orbits (which are spheres
centered at the origin) are labelled by the value of the Casimir
$K$.

For the CH top (\ref{VT}) the coadjoint orbits are embedded in the
Virasoro algebra (parameterized by the functions $L_k$) due to the
Lie-Poisson bracket (\ref{eq8}).

\section{Lax representation for the discrete Camassa-Holm equation and integrals of motion}

The Lax pair for the discrete CH equation (\ref{VT}) can be
obtained from the Lax pair for (\ref{CH}), \b \label{eq3}
\Psi_{xx}&=&\Big(\frac{1}{4}+\lambda (m+\frac{a}{2})\Big)\Psi
 \\\label{eq4}
\Psi_{t}&=&\Big(\frac{1}{2\lambda}-u+a\Big)\Psi_{x}+\frac{u_{x}}{2}\Psi,
\e as follows. We take the expansions \b \Psi&=&\sum_{n\in
\mathbb{Z}}\Psi_{\frac{n}{2}}e^{i\frac{n}{2}x}, \label{exp-Psi}\\
u&=&\frac{1}{2\pi}\sum_{n\in\mathbb{Z}}u_n e^{inx}+\frac{a}{2},
\qquad u_n=\frac{L_n}{1+n^2}\label{exp-u}. \e

\n Then (\ref{eq3}) gives \b
\frac{1}{\lambda}\Psi_{\frac{n}{2}}=\sum_{p\in\mathbb{Z}}\mathcal{L}_{\frac{n}{2},\frac{n}{2}-p}\Psi_{\frac{n}{2}-p},
\label{LaxMatrL}\e where \b
\mathcal{L}_{\frac{n}{2},\frac{n}{2}-p}=
-\frac{4}{n^2+1}\Big(\frac{L_p}{2\pi}+a\delta_{p,0}\Big),
\nonumber \e or \b \mathcal{L}_{\frac{n}{2}-q,\frac{n}{2}-p}=
-\frac{4}{(n-2q)^2+1}\Big(\frac{L_{p-q}}{2\pi}+a\delta_{p,q}\Big)\e

\n Now from (\ref{eq4}), (\ref{exp-Psi}), (\ref{exp-u})  and
(\ref{LaxMatrL}) it follows \b
\dot{\Psi}_{\frac{n}{2}}=\sum_{p\in\mathbb{Z}}\mathcal{A}_{\frac{n}{2},\frac{n}{2}-p}\Psi_{\frac{n}{2}-p},
\label{dotPsi}\e where \b
\mathcal{A}_{\frac{n}{2},\frac{n}{2}-p}&=&
-\frac{i}{4\pi}\Big(2n\frac{p^2+1}{n^2+1}+n-3p
\Big)u_p+in\Big(\frac{1}{4}-\frac{1}{n^2+1}\Big)a\delta_{p,0},
\nonumber \e or \b \mathcal{A}_{\frac{n}{2}-q,\frac{n}{2}-p}&=&
-\frac{i}{4\pi}\Big(2(n-2q)\frac{(p-q)^2+1}{(n-2q)^2+1}+n-3p+q
\Big)u_{p-q}\nonumber \\
&+&i(n-2q)\Big(\frac{1}{4}-\frac{1}{(n-2q)^2+1}\Big)a\delta_{p,q}.
\label{LaxMatrA}\e

\n Differentiating (\ref{LaxMatrL}) with respect to $t$ we obtain
\b\frac{1}{\lambda}\dot{\Psi}_{\frac{n}{2}}=\sum_{p\in\mathbb{Z}}\dot{\mathcal{L}}_{\frac{n}{2},\frac{n}{2}-p}\Psi_{\frac{n}{2}-p}+
\sum_{p\in\mathbb{Z}}\mathcal{L}_{\frac{n}{2},\frac{n}{2}-p}\dot{\Psi}_{\frac{n}{2}-p},
\nonumber \e

\n and with the further substitution from (\ref{dotPsi}),
\b\frac{1}{\lambda}\sum_{q\in\mathbb{Z}}\mathcal{A}_{\frac{n}{2},\frac{n}{2}-q}\Psi_{\frac{n}{2}-q}=
\sum_{p\in\mathbb{Z}}\dot{\mathcal{L}}_{\frac{n}{2},\frac{n}{2}-p}\Psi_{\frac{n}{2}-p}+
\sum_{p,q\in\mathbb{Z}}\mathcal{L}_{\frac{n}{2},\frac{n}{2}-q}\mathcal{A}_{\frac{n}{2}-q,\frac{n}{2}-p}\Psi_{\frac{n}{2}-p},
\nonumber \\
\sum_{q\in\mathbb{Z}}\mathcal{A}_{\frac{n}{2},\frac{n}{2}-q}\Big(\frac{1}{\lambda}\Psi_{\frac{n}{2}-q}\Big)=
\sum_{p\in\mathbb{Z}}\dot{\mathcal{L}}_{\frac{n}{2},\frac{n}{2}-p}\Psi_{\frac{n}{2}-p}+
\sum_{p,q\in\mathbb{Z}}\mathcal{L}_{\frac{n}{2},\frac{n}{2}-q}\mathcal{A}_{\frac{n}{2}-q,\frac{n}{2}-p}\Psi_{\frac{n}{2}-p},
\nonumber \e

\n and finally, the substitution of (\ref{LaxMatrL}) gives
\b\sum_{p,q\in\mathbb{Z}}\mathcal{A}_{\frac{n}{2},\frac{n}{2}-q}\mathcal{L}_{\frac{n}{2}-q,\frac{n}{2}-p}\Psi_{\frac{n}{2}-p}=\phantom{*************}
\nonumber \\
\sum_{p\in\mathbb{Z}}\dot{\mathcal{L}}_{\frac{n}{2},\frac{n}{2}-p}\Psi_{\frac{n}{2}-p}+
\sum_{p,q\in\mathbb{Z}}\mathcal{L}_{\frac{n}{2},\frac{n}{2}-q}\mathcal{A}_{\frac{n}{2}-q,\frac{n}{2}-p}\Psi_{\frac{n}{2}-p},
\label{LaxMatrForm} \e

\n or in matrix form, \b \dot{\mathcal{L}}=[\mathcal{A},
\mathcal{L}]. \label{LaxMatrF2} \e

\n After some lengthy computations one can verify that
(\ref{LaxMatrF2}) gives (\ref{VT}). The integrals of motion are
given by $I_k=\text{tr}(\mathcal{L}^k)$. For example, \b
I_1=\text{tr}(\mathcal{L})=\sum_{p\in\mathbb{Z}}\mathcal{L}_{\frac{n}{2}-p,\frac{n}{2}-p}=
-4\Big(\frac{L_{0}}{2\pi}+a\Big)\sum_{p\in\mathbb{Z}}\frac{1}{(n-2p)^2+1}
\nonumber\e

\n produces, up to an overall constant, the Casimir $H_0$,
(\ref{H0}). \b
I_2&=&\text{tr}(\mathcal{L}^2)=\sum_{p,q\in\mathbb{Z}}\mathcal{L}_{\frac{n}{2}-p,\frac{n}{2}-q}\mathcal{L}_{\frac{n}{2}-q,\frac{n}{2}-p}
\nonumber \\
&=&
\frac{4}{\pi^2}\sum_{p,q\in\mathbb{Z}}\frac{L_{p-q}L_{q-p}}{[(n-2p)^2+1][(n-2q)^2+1]}\nonumber
\\ &+&\frac{16a}{\pi}(L_0+\pi
a)\sum_{p\in\mathbb{Z}}\frac{1}{[(n-2p)^2+1]^2}.\e

\n With partial fractions decomposition with respect to $n$ one
can derive the identity \b
\frac{1}{[(n-2p)^2+1][(n-2q)^2+1]}\phantom{************************}\nonumber
\\ =\frac{1/4}{(p-q)^2+1}\Big\{\frac{(n-2q)+(p-q)}{(p-q)[(n-2q)^2+1]}-\frac{(n-2p)+(q-p)}{(p-q)[(n-2p)^2+1]}\Big\}. \nonumber\e
\n Further, using the fact that all expressions that change sign
under $p-q \rightarrow -(p-q)$ are zero, due to the summation over
all integer numbers, we have \b
\frac{4}{\pi^2}\sum_{p,q\in\mathbb{Z}}\frac{L_{p-q}L_{q-p}}{[(n-2p)^2+1][(n-2q)^2+1]}\phantom{*********************}\nonumber
\\= \frac{1}{\pi^2}\sum_{p,q\in\mathbb{Z}}\frac{L_{p-q}L_{q-p}}{1+(p-q)^2}\Big\{\frac{1}{(n-2p)^2+1}+\frac{1}{(n-2q)^2+1}\Big\}\nonumber \\
=\frac{2}{\pi^2}\sum_{p\in\mathbb{Z}}\frac{L_{p}L_{-p}}{1+p^2}\sum_{q\in\mathbb{Z}}\frac{1}{(n-2q)^2+1}.
\nonumber \e Thus, the new integral that appears is
$\sum_{p\in\mathbb{Z}}\frac{L_{p}L_{-p}}{1+p^2}$, giving $H_1$,
the first Hamiltonian (\ref{ham}).

\section{Oscillator algebra, Miura transformation and momentum map}

Let us introduce now the oscillator algebra \b
i\{a_n,a_m\}=\frac{2\pi a}{\kappa^{2}}n\delta_{n+m,0}, \label{oa}
\e

\n where $\kappa$ is an arbitrary constant. Clearly, $a_0$ is a
Casimir due to (\ref{oa}). One can easily verify the following
oscillator representation of the Virasoro algebra \cite{K85,HK90}:
\b L_n=-\kappa(n-1)a_n+\frac{\kappa^2}{4\pi a}\sum_{k\in
\mathbb{Z}}a_k a_{n-k}. \label{oscrep}\e

\n This representation is also known as Sugawara construction.
Further, it is evident that \b i\{a_n,L_m\}=na_{n+m}+\frac{2\pi
a}{\kappa}n(n+1)\delta_{n+m,0}. \label{aLcom} \e

Since $a_k$ satisfy the 'canonical' Poisson brackets they are
natural candidates for the coordinates in the phase-space. Thus,
$L_n$ has an interpretation of a momentum and (\ref{oscrep}) gives
the momentum map. The Sugawara construction relates to the Miura
transformation, which in terms of field variables can be obtain as
follows. Defining \b v=\frac{1}{2\pi}\sum_{k\in \mathbb{Z}}a_k
e^{ikx}+\frac{a}{\kappa} \e from (\ref{oscrep}) and (\ref{exp-m})
we have the analog of the Miura transformation: \b m=i\kappa
v_x+\frac{\kappa^2}{2a}v^2+\frac{a}{2} \label{mi} \e

\n The reality can be achieved by taking $\kappa$ purely
imaginary, $a_k=\bar{a}_{-k}$ for $k\ne 0$ and $\kappa=2\pi i
a/\Im (a_0)$.

\n Here we notice that the Casimir (\ref{Cas}) due to (\ref{mi})
leads to the restriction \b \int_{0} ^{2 \pi} v^2(x,t) {\text
d}x=\text{const}, \e

\n which reduces the evolution of $v(x,t)$ on the $L^2$-sphere. In
terms of the canonical coordinates this condition is \b
\sum_{k>0}|a_k|^2=\text{const}, \label{l2sphere} \e since $a_0$ is
a constant. It shows that the time evolution of the canonical
variables, given by \b \dot{a}_n=\{a_n,H_1\}  \nonumber \e is
constrained on the infinite-dimensional $l_2$-sphere, a condition,
similar to the one that we see in the $so(3)$ example (\ref{K}).

When $a=0$, the Sugawara construction for the Virasoro modes in
the case of zero central charge is \b
L_n=\frac{1}{2\tilde{\kappa}}\sum_{k\in \mathbb{Z}}a_ka_{n-k},
\nonumber \e where $\tilde{\kappa}$ is an arbitrary constant and
\b i\{a_n,a_m\}=\tilde{\kappa} n\delta_{n+m,0}. \label{alg0} \e

\n The Casimir with respect to the first Poisson bracket
(\ref{PB_1}) with $a\equiv 0$ is \b \int_{0}^{2\pi} \sqrt{m}
\text{d}x=\sqrt{\frac{\pi}{\tilde{\kappa}}}\int_{0}^{2\pi} v
\text{d}x=2\pi\sqrt{\frac{\pi}{\tilde{\kappa}}} a_0, \nonumber \e

\n i.e. this is the Casimir $a_0$ of (\ref{alg0}).

\n With the expansions \b m
=\frac{1}{2\pi}\sum_{n\in\mathbb{Z}}L_n e^{inx}, \qquad
v=\frac{1}{2\pi}\sum_{n\in\mathbb{Z}}a_n e^{inx}\nonumber \e the
Sugawara construction takes the form  $ m
=\frac{\pi}{\tilde{\kappa}} v^{2}. $ Since the integral
$\int_{0}^{2\pi} m \text{d}x = \text{const}$ is a Casimir, we have
again \b \int_{0}^{2\pi} v^2 \text{d}x=\text{const}, \nonumber\e
leading to (\ref{l2sphere}).

\section{The Hunter-Saxton equation}

The Hunter-Saxton (HS) equation \b u_{xxt}+2u_xu_{xx}+uu_{xxx}=0
\nonumber  \e describes the propagation of waves in a massive
director field of a nematic liquid crystal \cite{HS}, with the
orientation of the molecules described by the field of unit 1
vectors $n(x,t) =(\cos u(x,t), \sin u(x,t))$, where $x$ is the
space variable in a reference frame moving with the linearized
wave velocity, and $t$ is a 'slow time variable'. A linear term
$au_{xxx}$ can be generated by a shift $u\rightarrow u+a$: \b
u_{xxt}+au_{xxx}+2u_xu_{xx}+uu_{xxx}=0. \label{HS} \e The HS
equation is a short-wave limit of the CH equation, and can be
obtained if one takes $m=-u_{xx}$. The Hamiltonian representation
(\ref{CHPB}) -- (\ref{H_1}) for this equation is also valid. The
HS equation (\ref{HS}) is an integrable, bi-Hamiltonian equation
with a second Hamiltonian representation $ m_t=\{m,H_2\}_2$, where
\b \{F,G\}_2 &\equiv& \int_{0} ^{2 \pi}\frac{\delta F}{\delta
m}\partial^3\frac{\delta G}{\delta m}{\text d}x, \label{HSPB_2}\\
H_2&=&\frac{1}{2}\int_{0} ^{2 \pi} (u-\frac{a}{2})u_x^2 {\text
d}x. \label{HSH_2} \e

\n The HS Lax pair is \b \label{eq30} \Psi_{xx}&=&\lambda m\Psi,
 \\\label{eq40}
\Psi_{t}&=&\Big(\frac{1}{2\lambda}-u-a\Big)\Psi_{x}+\frac{u_{x}}{2}\Psi.
\e

The analytic and geometric aspects of the HS equation are
discussed in a variety of recent papers, e.g. \cite{HZ,BC05,L07}
and the references therein.

\n Again, assuming periodicity and using the expansions \b
\Psi=\sum_{n\in \mathbb{Z}}\Psi_{n}e^{inx}, \qquad
u=\frac{1}{2\pi}\sum_{n\in\mathbb{Z}}u_n e^{inx}\nonumber \e we
obtain the discrete form of the HS equation: \b
in\dot{u}_n-an^2u_n-\frac{1}{2}\sum_{k\in
\mathbb{Z}}k(n+k)u_ku_{n-k}=0.\nonumber \e

\n In a similar manner from the Lax pair we obtain the matrix Lax
representation for the discrete HS equation \b
\dot{\mathcal{L}}^{HS}=[\mathcal{A}^{HS}, \mathcal{L}^{HS}].
\label{LaxMatrF2-1} \e where \b
\mathcal{L}^{HS}_{n,n-p}=-\frac{p^2}{n^2}u_p, \qquad
\mathcal{A}^{HS}_{n,n-p}=\frac{i}{2}\Big(-\frac{p^2}{n}-2n+3p
\Big)u_p -ina\delta_{p,0}.\nonumber \e

The momentum map (the Sugawara construction) for the HS equation
remains the same as for the CH equation. However, it becomes
degenerated in the case $a=0$, since $m=-u_{xx}$ and the Casimir
$\int_{0}^{2\pi} m \text{d}x = 0$. Then $ \int_{0}^{2\pi} v^2
\text{d}x=0$, which, for real variables is only possible when
$v\equiv 0$, i.e. $m\equiv0$.

\section{Conclusions}

At the examples of the CH and HS equations we have shown that the
integrable systems with quadratic Hamiltonians are equivalent to
integrable tops (possibly with infinitely many components),
associated to the algebra of their Poisson brackets. An example
for the two dimensional Euler equations in fluid mechanics is
presented in \cite{Z91}, another example for the KdV superequation
in \cite{K85,OK87}.

\section*{Acknowledgments}

The support of the G. Gustafsson Foundation for Research in
Natural Sciences and Medicine is gratefully acknowledged. The
author is thankful to Prof. A. Constantin and Dr G. Grahovski for
many stimulating discussions.

\label{lastpage}
\end{document}